\documentclass[prb,aps,twocolumn,showpacs]{revtex4-1}
\usepackage{amsfonts,amsmath,bm,multirow}
\usepackage[OT4]{fontenc}

\usepackage{graphicx}
\usepackage{xcolor}
\usepackage{epstopdf}
\usepackage{dsfont}
\usepackage{tabularx}
\usepackage{hf-tikz}

\usepackage{physics}
\usepackage{hyperref}
\usepackage{placeins}

\newcommand{\kp}{\bm{k}\!\vdot\!\bm{p}}
\newcommand{\rr}{\bm{r}}

\begin{document}

\title{Spin-orbit coupling and spin relaxation of hole states in [001]- and [111]-oriented quantum dots of various geometry}
\author{Krzysztof Gawarecki}
\email{Krzysztof.Gawarecki@pwr.edu.pl} 
\author{Mateusz Krzykowski}
\affiliation{Department of Theoretical Physics, Faculty of Fundamental Problems of Technology, Wroc\l{}aw University of Technology, Wybrze\.ze Wyspia\'nskiego 27, 50-370 Wroc\l{}aw, Poland}

\begin{abstract}
	We study the influence of spin-orbit coupling on the hole states in InAs/GaAs quantum dots grown on [$001$]- and [$111$]-oriented substrates belonging to symmetry point groups: $C_{\mathrm{2v}}$, $C_{\mathrm{3v}}$ and $D_{\mathrm{2d}}$. We investigate the impact of various spin-orbit mechanisms on the strength of coupling between $s$- and $p$-shell states, which is a significant spin-flip channel in quantum dots. We calculate spin relaxation rates between the states of lowest Zeeman doublet and show that the [$111$]-oriented structure offers one order of magnitude slower relaxation compared to the usual [$001$]-oriented self-assembled QD. 
	The magnetic-field dependence of the hole states is calculated using multiband (up to $14$ bands) $\kp$ model.  We identify the irreducible representations linked to the states and discuss the selection rules, which govern the avoided-crossing pattern in magnetic-field dependence of the energy levels.  We show that dominant contribution to the coupling between some of these states comes from the shear strain.  On the other hand, we demonstrate no coupling between s- and p-shell states in the [$111$]-oriented structure. 
\end{abstract}

\maketitle

\section{Introduction}
\label{sec:intr}

The properties of nanostructures related to the spin degree of freedom are interesting from the point of view of possible applications in quantum information proccesing and spintronics\cite{Zutic2004,Hanson2007,loss98,yong13}. Coupling of spin to orbital degrees of freedom via the spin-orbit coupling (SOC) influences the carrier spectrum and could provide a channel of quantum coherent spin control\cite{Flindt2006}. On the other hand, it may mix spin configurations, which leads to spin relaxation and dephasing processes\cite{Khaetskii2000a,Khaetskii2000b,Golovach2008,Climente2013,Mielnik-Pyszczorski2018b,Segarra2015a}. 
The lack of inversion symmetry, on the level of crystal lattice (bulk inversion asymmetry, BIA), in the shape of a nanostructure, or induced by external electric field (structure inversion asymmetry, SIA) gives rise to  Dresselhaus and Rashba spin-orbit coupling, respectively\cite{Winkler2003}. Furthermore, recent investigations show hidden spin polarization in centrosymmetric crystals\cite{Zhang2014}.

Dresselhaus and/or Rashba interactions are commonly accounted for theoretically within effective models\cite{Bulaev2005,Siranush12,Siranush13,Manaselyan09}. While the parameters are available and well established for bulk materials, in the case of nanostructures the coupling strength is determined by their shape, composition profile, substrate orientation, strain and abrupt material interfaces. In consequence, a reliable quantitative description of the SIA effects requires advanced modeling. The impact of Dresselhaus and Rashba couplings on carrier states in a quantum dot (QD) were studied in various approaches\cite{Bulaev2005,Siranush12,Siranush13,Gawarecki2018a,Manaselyan09}. In~the case of InAs/GaAs self-assembled QDs, the influence of spin-orbit coupling is more complicated due to the presence of interfaces abrupt and symmetry-breaking shear strain. The latter gives spin admixture leading to electron spin relaxation\cite{Mielnik-Pyszczorski2018a} and this is one of the most important factors determining the splitting between hole $p$-type states\cite{Gawarecki2018a}. Symmetry of the self-assembled QD plays crucial role for its optical properties and exchange interaction\cite{Zielinski2015a,Ehrhardt2014}. It has been also shown that the coupling between $s$- and $p$-shell electron states related to the Rashba interaction is enhanced by the dot anisotropy\cite{Siranush12,Siranush13}.  Furthermore, the symmetry determines anticrossing pattern\cite{Doty2010,daniels13,Ardelt2016} as well as affects spin mixing and relaxation in a double QD system\cite{Segarra2015a,Ma2016}. The properties of the nanostructure depends not only on its geometrical shape but also orientation of the underlying substrate. Due to potential application for entangled photon pairs generation, [$111$]-oriented QDs were subject of many theoretical and experimental works\cite{Singh2009, Schulz2008, Stock2010, Mano2010, Ostapenko2010, schultz11, Marquardt2014,Swiderski2017}.

The phonon-induced spin-flip of carriers in QDs was investigated in many theoretical\cite{Bulaev2005,Lu2005,Bulaev2005b,woods04,Mielnik-Pyszczorski2018b} and experimental works\cite{heiss07,Kroutvar2004,Scarlino2014a}. This effect results from the direct spin-phonon coupling\cite{Roth1960} and from spin-admixture mechanisms\cite{Khaetskii2000a,Khaetskii2000b}. The latter is related to the fact, that in presence of the spin-orbit interaction, the state with a given nominal spin orientation, contains a non-zero component of opposite spin. Due to this admixture, even diagonal part of the carrier-phonon Hamiltonian gives rise to the spin-flip effect. In fact, it has been shown for the electron and hole in 2D GaAs QD, that near the avoided crossing between $s$ and $p$ states (where $p$-type spin admixture to the $s$-type state becomes large) the spin-flip transition rate dramatically increases\cite{Bulaev2005,Bulaev2005b}. Also, the orientation of magnetic field with respect to crystallographic axes strongly affects spin relaxation time in gate-defined GaAs QDs, which was attributed to the interplay of Rashba and Dresselhaus couplings\cite{Scarlino2014a}. This was addressed theoretically for [$001$]- and [$111$]-grown GaAs cuboidal QDs\cite{Segarra2015}. Furthermore, in Ref.~\onlinecite{Lu2005} hole spin relaxation is studied for a QD defined by a parabolic potential in GaAs quantum well (QW) of various crystal orientation. The influence of strain on spin-flip transitions was investigated. It has been shown, that for [$001$]-grown structure, the biaxial strain affects spin relaxation indirectly, by changing the energy difference between heavy- and light-hole subbands. On the other hand, for QD in [$111$]-oriented QW, strain provides a direct spin mixing channel. Furthermore, such a structure, offers longer spin relaxation time (compared to the [$001$]-oriented one) at the regime of strong confinement in growth direction\cite{Lu2005}.

In this work, we investigate the influence of various mechanisms (Dresselhaus and Rashba interaction, shear strain) on the coupling between the $s$- and $p$-shell hole states in InAs/GaAs QDs. Within the $8$-band $\kp$ model, we calculate the magnetic-field dependence of the energy levels and study the width of avoided crossing between the $s$- and $p$-type state.  We take into account [$001$]- and [$111$]-oriented substrates and consider three types of QDs representing $C_{\mathrm{2v}}$, $C_{\mathrm{3v}}$ and $D_{\mathrm{2d}}$ symmetry point groups. For these, we identify the irreducible representations of hole states, discuss the selection rules, and demonstrate the absence of coupling between $s$ and both $p$-shell states for the [$111$]-oriented structure. Finally, we show that spin-flip transitions at low magnetic fields are slower by an one order of magnitude for [$111$]-oriented QD compared to the standard [$001$]-oriented system.

The paper is organized as follows. In Sec.~\ref{sec:model}, the methods used to calculate the strain distribution and the carrier states are described.  In Sec.~\ref{sec:results}, we present the results of numerical simulations for all of the considered structures. The phonon-induced spin relaxation is discussed in Sec.~\ref{sec:phonon}. Finally, Sec.~\ref{sec:conclusions} contains the summary. In Appendix A we present character tables of the symmetry point groups used in the paper. In Appendix B, we describe the effective model with empirical parameters which are fitted to the numerical data.

\section{Model}
\label{sec:model}


\begin{figure}[tb]
	\begin{center}
		\includegraphics[width=90mm]{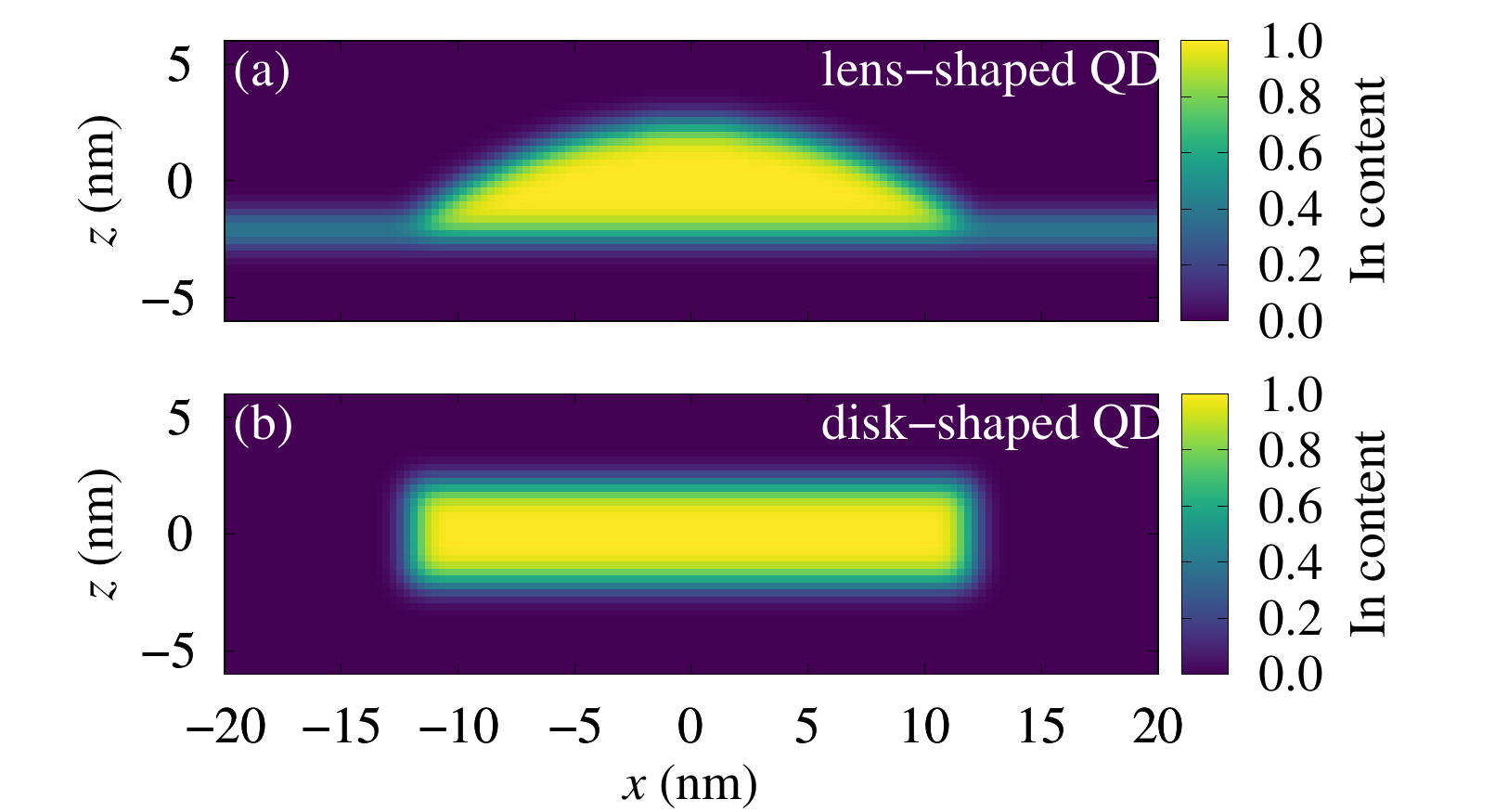}
	\end{center}
	\caption{\label{fig:comp}\textcolor{gray}({Color online) Material distribution in the system, in the case of lens- (a) and disk-shaped (b) QD. } }
\end{figure}
The system under consideration contains a single InAs/GaAs QD. We model lens- and disk-shaped QDs [see Fig.~\ref{fig:comp}(a,b)].
In both cases, the dot height is $h=4.2$~nm and the base radius is $r=12$~nm. Furthermore, the lens-shaped dot is placed on a $0.6$~nm thick wetting layer.

The distribution of strain in the system is calculated within the continuous elasticity approach\cite{pryor98b}. 
To calculate the strain tensor elements for the [$111$]-grown system, we perform transformation to the rotated coordinates\cite{schultz11}. 
The piezoelectric potential is calculated up to the second order in the strain tensor elements\cite{Bester06b} with parameters taken from Ref.~\onlinecite{Caro2015}, while transformation to the [$111$]-oriented system is performed following Ref.~\onlinecite{schultz11}.

The hole states are calculated using $8$- and $14$-band $\kp$ method in the envelope function approximation (if~not stated otherwise, we take into account $8$ bands). 
The full $14$-band Hamiltonian can be divided into the blocks corresponding to the irreducible representations of $\mathrm{T_{d}}$ symmetry point group: $\Gamma_{8c}$, $\Gamma_{7c}$, $\Gamma_{6c}$, $\Gamma_{8v}$, or $\Gamma_{7v}$\cite{Winkler2003,eissfeller12,Gawarecki2018a}
\[H=
\left(
\begin{array}{*{5}{c}}
H_{\mathrm{8c8c}} & H_{\mathrm{8c7c}} & H_{\mathrm{8c6c}} & H_{\mathrm{8c8v}} & H_{\mathrm{8c7v}}\\
H_{\mathrm{7c8c}} & H_{\mathrm{7c7c}} & H_{\mathrm{7c6c}} & H_{\mathrm{7c8v}} & H_{\mathrm{7c7v}}\\
H_{\mathrm{6c8c}} & H_{\mathrm{6c7c}} & \tikzmarkin{a}(0.05,-0.09)(-0.15,0.31) H_{\mathrm{6c6c}} & H_{\mathrm{6c8v}} & H_{\mathrm{6c7v}}\\
H_{\mathrm{8v8c}} & H_{\mathrm{8v7c}} & H_{\mathrm{8v6c}} & H_{\mathrm{8v8v}} & H_{\mathrm{8v7v}}\\
H_{\mathrm{7v8c}} & H_{\mathrm{7v7c}} & H_{\mathrm{7v6c}} & H_{\mathrm{7v8v}} & H_{\mathrm{7v7v}} \tikzmarkend{a} 
\end{array}
\right),
\]
where highlighted part corresponds to the standard $8$-band $\kp$ Hamiltonian.  The states are then $8$- or $14$-component pseudo-spinors, where each part refers to one of the subbands.
The full Hamiltonian, parameters and details of numerical implementation are presented in the Appendix of Ref.~\onlinecite{Gawarecki2018a}. We account for an axial electric field by adding a diagonal term $H^{\mathrm{(Efield)}} = \abs{e} F z$ to the Hamiltonian, where $e$ is the elementary charge and $F$ is the field magnitude.
To model [$111$]-oriented system, we rotate the Hamiltonian transforming all vectors and invariant matrices (see a detailed description in Ref.~\onlinecite{eissfeller12}).
The average values of the $z$ projection of the hole envelope angular momenta are calculated from
\begin{align*}
\langle M_{z} \rangle &= \sum_{m=1}^{8} \int_{-\infty}^{\infty} \psi^{*}_{n,m}(\rr)  \left( i y \frac{\partial}{\partial x} - i x  \frac{\partial}{\partial y}  \right)  \psi_{n,m}(\rr) \mathrm{d} \rr.
\end{align*} 
where $\psi_{n,m}(\rr)$ denotes the $m$-th band component of the $n$-th hole wave function.


\begin{figure*}[!tb]
	\begin{center}
		\includegraphics[width=7.0in]{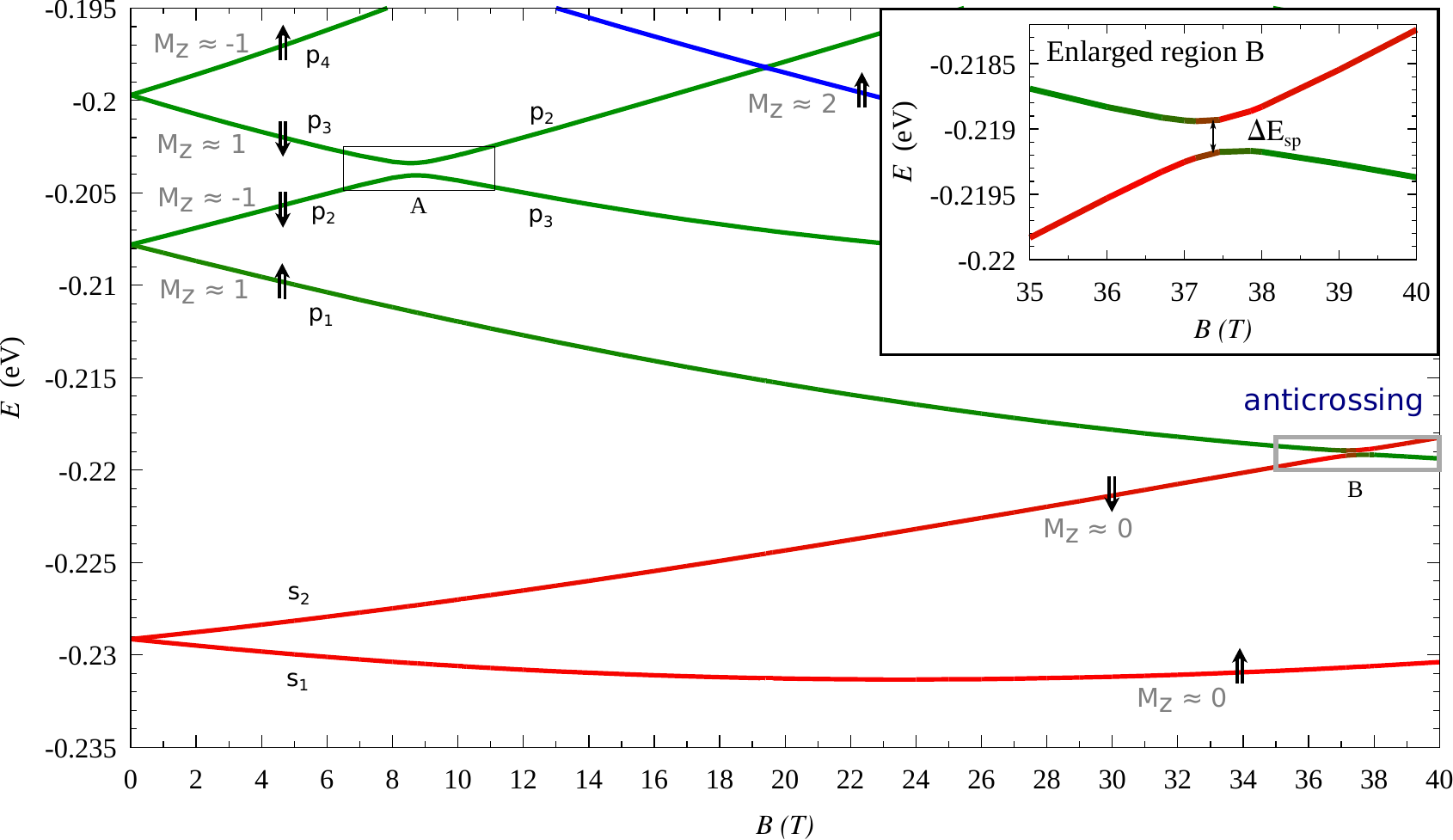}
	\end{center}
	\caption{\label{fig:mag_h}\textcolor{gray}({Color online) Magnetic-field dependence of the lowest hole energy levels for the lens shaped [001]-oriented QD. The inset contains enlarged part of the plot with anticrossing between $s$- and $p$-type states. Energy $E=0$ refers to the unstrained GaAs valence-band edge. } }
\end{figure*}

\begin{figure*}[!tb]
	\includegraphics[width=7in]{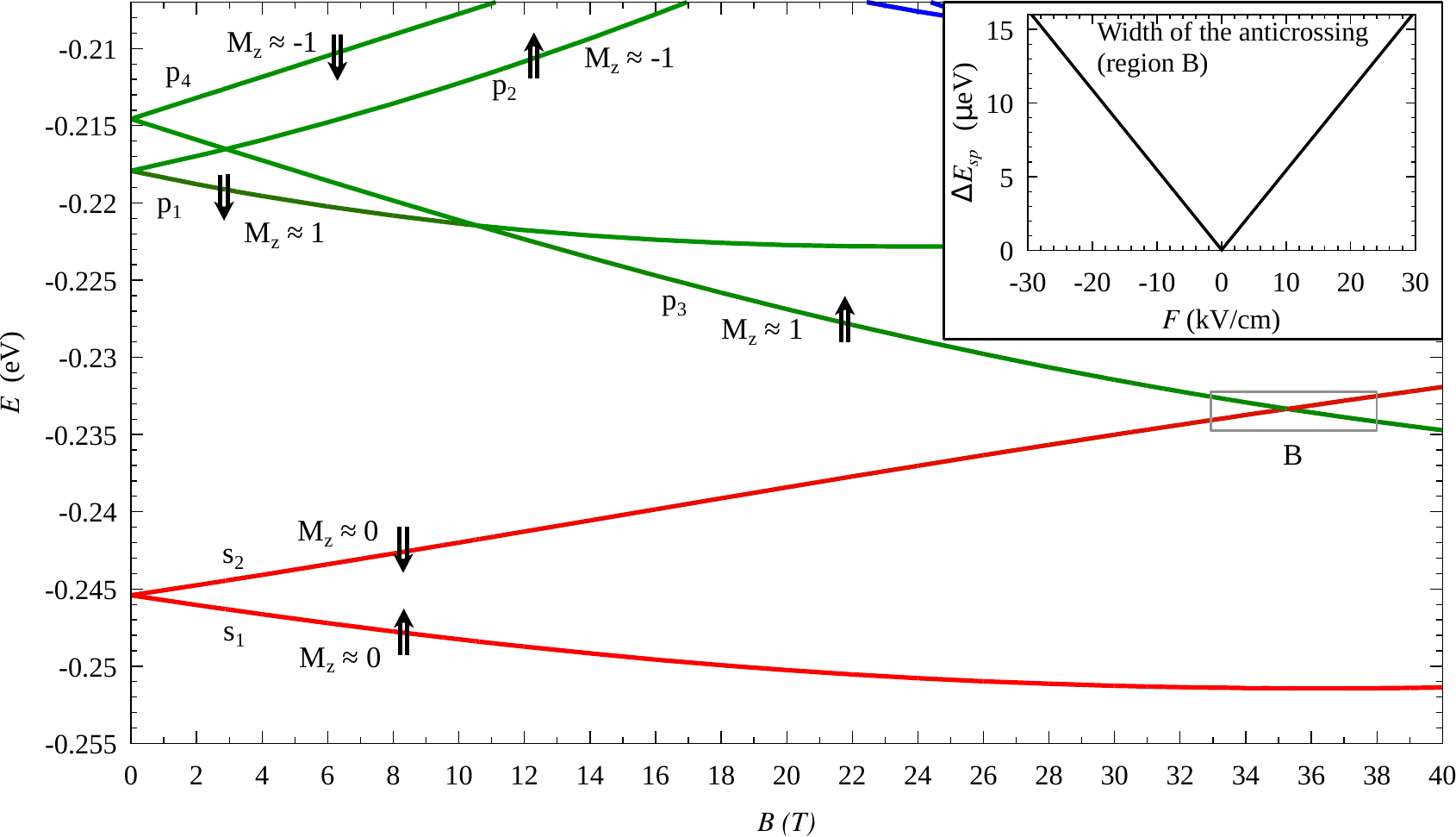}
	\caption{\label{fig:cqd}\textcolor{gray}({Color online) Magnetic-field dependence of the lowest hole energy levels for the disk shaped [001]-oriented QD. The inset contains the avoided crossing width between $s$- and $p$-shell energy levels as a function of external axial electric field $F$. } }
\end{figure*}

The $14$-band $\kp$ model accounts  inherently for the Dresselhaus and Rashba couplings\cite{Winkler2003}. Also the $8$-band model contains the most important terms for the Rashba coupling, while the Dresselhaus interaction is described by perturbative elements explicitly added to $H_{\mathrm{6c8v}}$ and $H_{\mathrm{6c7v}}$.
The Dresselhaus SOC Hamiltonian for the electron (in $2$-band $\kp$ model) can be approximated by
$ H^{\mathrm{(D)}}_{\mathrm{6c6c}} \propto \langle k^{2}_{z} \rangle \qty ( k_{+} \sigma_{+} + k_{-} \sigma_{-} )$, where $k_{\pm} = k_{x} \pm i k_{y}$, and  $\sigma_{\pm}$ is the spin ladder operator. This couples $\ket{M_{z} \approx 0, \downarrow}$ to $\ket{M_{z} \approx 1, \uparrow}$ and $\ket{M_{z} \approx 0, \uparrow}$ to $\ket{M_{z} \approx -1, \downarrow}$, where $\uparrow, \downarrow$ refers to the spin orientation. In contrast, the Rashba coupling approximated by $ H^{\mathrm{(R)}}_{6c6c} \propto i \qty ( k_{+} \sigma_{-} - k_{-} \sigma_{+} )$ connects $\ket{M_{z} \approx 0, \uparrow}$ to $\ket{M_{z} \approx 1, \downarrow}$, and $\ket{M_{z} \approx 0, \downarrow}$ to $\ket{M_{z} \approx -1, \uparrow}$.
On the other hand, the influence of the spin-orbit interaction for holes is much more complicated compared to the electron\cite{Winkler2003}. In this case, the Rashba coupling may mix $\ket{M_{z} \approx 0, \Uparrow}$ to both $\ket{M_{z} \approx 1, \Downarrow}$ and $\ket{M_{z} \approx -1, \Downarrow}$ (and vice-versa), where $\Uparrow,\Downarrow$ refers to band angular momentum (see Appendix B). 

\section{Numerical results and symmetry classification}
\label{sec:results}
\subsection{[001]-oriented lens shaped QD }
We calculated the magnetic-field dependence of the lowest-energy hole states in the lens-shaped QD. The shape of such a structure does not have the inversion symmetry. The energy levels obtained from $8$-band $\kp$ simulations are presented in Fig.~\ref{fig:mag_h}. Although the states contain contributions of various envelope symmetry (which results from the subband mixing), they can be labeled as $s$, $p$, $d$, ... with respect to the dominant component (read from the value of $\langle M_{z} \rangle$).  The two lowest-energy states (marked by the red lines) exhibit $s$-type symmetry, their average value of the axial projection of envelope angular momentum $\langle M_{z} \rangle$ is close to 0. The next four states (plotted with green lines) exhibit $p$-type symmetry with $\langle M_{z} \rangle \approx \pm 1$. 
Although the shape of the QD transforms according to the $C_{\infty v}$ group, the underlying crystal lattice limits the symmetry of the system to the $C_{\mathrm{2v}}$ (at $B=0$~T).  Due to the spin-oribit coupling, the system must be described in terms of the double group representations\cite{Dresselhaus2010,Yu2010,Bir1974}.  The symmetry point group $C_{\mathrm{2v}}$ contains only one irreducible double group representation $D_{1/2}$ (see the character table in Appendix A) and all states must belong to it. Since $D_{1/2}$ is two-dimensional, the states are doubly degenerate (which in fact results from the time-reversal symmetry). At nonzero axial magnetic field $B \neq 0$, the symmetry of the system is further reduced to $C_{2}$\cite{Winkler2003,Segarra2015a}. In this case, $D_{1/2}$ splits into two one-dimensional representations: $D_{A}$ and $D_{B}$, where $D_{A} = D^{*}_{B}$ (see Tab.~\ref{tab:c2} in Appendix A). For each state~$\ket{\Psi}$ we found the relevant irreducible representation $\alpha$ via projection $\widehat{P}^{(\alpha)} \ket{\Psi}$, where $\widehat{P}^{(\alpha)} = \sum_{i}  \chi^{*} (\widehat{R}_{i}) \widehat{R}_{i}$, and $\chi (\widehat{R}_{i}) $ is the character of the representation $\alpha$ for the symmetry operation $\widehat{R}_{i}$\cite{Dresselhaus2010,Piela}.  The states $s_{1}$, $p_{2}$ and $p_{3}$ belong to $D_{A}$, whereas $s_{2}$, $p_{1}$ and $p_{4}$ to $D_{B}$.  According to the selection rules, two states can couple if they belong to the same irreducible representation.  In the presence of SOC, an avoided crossing pattern appears in the system spectrum. In the considered system, the spin-orbit coupling in the hole $p$ shell favors the parallel orientation of the envelope and band angular momenta (see a detailed discussion in Ref.~\onlinecite{Gawarecki2018a}). At $B \approx 9$~T, there is an avoided crossing between $p_{2}$ and $p_{3}$ (region A in Fig.~\ref{fig:mag_h}), they have the same orientation of the band angular momenta but different $M_{z}$. 
Furthermore, an avoided crossing appears between states $s_{2}$ ($\langle M_{z} \rangle \approx 0, \Downarrow$) and $p_{1}$ ($ \langle M_{z} \rangle \approx 1, \Uparrow$) at region B, where its width is $\Delta E_{\mathrm{sp}}=0.246$~meV. 

To assess the importance of various SOC mechanisms and check the accuracy of $8$-band $\kp$, we compared the value of $\Delta E_{\mathrm{sp}}$  obtained within several degrees of approximation. As shown in Tab.~\ref{tab:compar}, the results from $8$- and $14$-band $\kp$ are in a good agreement. Dresselhaus terms are negligible for $\Delta E_{\mathrm{sp}}$, however they could be important for $s_{1}$ - $p_{3}$ and $s_{2}$ - $p_{4}$ couplings (which is hard to estimate, because it is not represented by any avoided crossing in the considered spectrum). In the last approach, the influence of shear strain in the valence band is neglected by setting the deformation potential $d_{\mathrm{v}} = 0$. In this case, $\Delta E_{\mathrm{sp}}$ is significantly reduced, which suggest that the shear strain is one of the most important factors determining the $s$ - $p$ coupling.
\begin{table}[h]
	\caption{\label{tab:compar}The anticrossing width $\Delta E_{\mathrm{sp}}$ between $s$ and $p$-type state obtained from various approximations.}
	\begin{ruledtabular}
		\begin{tabular}{lc}
			Model & $\Delta E_{\mathrm{sp}}$~(meV)  \\ 
			\hline  \\ [-1.5ex]
			14-band $\kp$, full  & $0.25481$ \\			 	
			8-band $\kp$, full  & $0.24627$ \\				 
			8-band $\kp$, Dresselhaus terms $H^{\mathrm{(D)}} = 0$  & \multirow{2}{*}{$0.24565$}  \\
			 neglected   &   \\			 
		  8-band $\kp$, shear strain neglected ($d_{\mathrm{v}} = 0$)  & $0.11679$ 
		\end{tabular} 
	\end{ruledtabular}
\end{table}

The Rashba coupling can rise due to external potentials. We calculated $\Delta E_{\mathrm{sp}}$ at the axial electric field $F = 30$~kV/cm and obtained $\Delta E_{\mathrm{sp}}=0.243$~meV, while opposite direction $F = -30$~kV/cm led to $\Delta E_{\mathrm{sp}}=0.248$~meV. This shows, that for the considered QD the axial electric field generates the Rashba coupling, which is much weaker than the structure inversion asymmetry resulting from the QD shape.

\subsection{[001]-oriented disk shaped QD}
The magnetic-field dependence of energy levels calculated for the [$001$]-oriented disk-shaped QD is presented in Fig.~\ref{fig:cqd}. For such a system, at $B=0$, the symmetry point group is $D_{\mathrm{2d}}$. According to the character table (Tab.~\ref{tab:d2d} in Appendix A), there are two irreducible double-group representations $D_{1/2}$ and $D'$. In the presence of magnetic field, the symmetry of the system is reduced to $S_{4}$ (see Tab.~\ref{tab:s4} in Appendix A)\cite{Winkler2003}. Then, the states $s_{1}$ and $p_{1}$ belong to $D_{\mathrm{I}}$,  $s_{2}$ and $p_{2}$ to $D_{\mathrm{II}}$, $p_{3}$ to $D_{\mathrm{IV}}$, and $p_{4}$ to $D_{\mathrm{III}}$ representation. Since $s_{2}$ and $p_{3}$ states belong to different representations, there is no avoided crossing between their energy levels (see region B in Fig.~\ref{fig:cqd}). For the same reason, we obtain a crossing between $p_{2}$ and $p_{3}$ at about $3$~T. In contrast to the lens-shaped QD, at weak magnetic field, the states with antiparallel envelope and band angular momenta have lower energy compared to the opposite configuration. 

The symmetry of the system can be further reduced by external electric field. For axial field, the symmetry changes from $D_{\mathrm{2d}}$ to $C_{\mathrm{2v}}$ (and from $S_{4}$ to $C_{2}$ at $B \neq 0$). In this case, the Rashba coupling between upper $s$-shell state ($M_{z} \approx 0$, $\Downarrow$) and the $p$-shell state ($M_{z} \approx 1$, $\Uparrow$) appears. The simulation results are presented in the inset of Fig.~\ref{fig:cqd}. The width of the anticrossing increases linearly with the electric field, and at $F=0$ there is a crossing between the relevant energy levels.

\subsection{[111]-oriented lens shaped QD }
\begin{figure}[!tb]
	\begin{center}
		\includegraphics[width=3.2in]{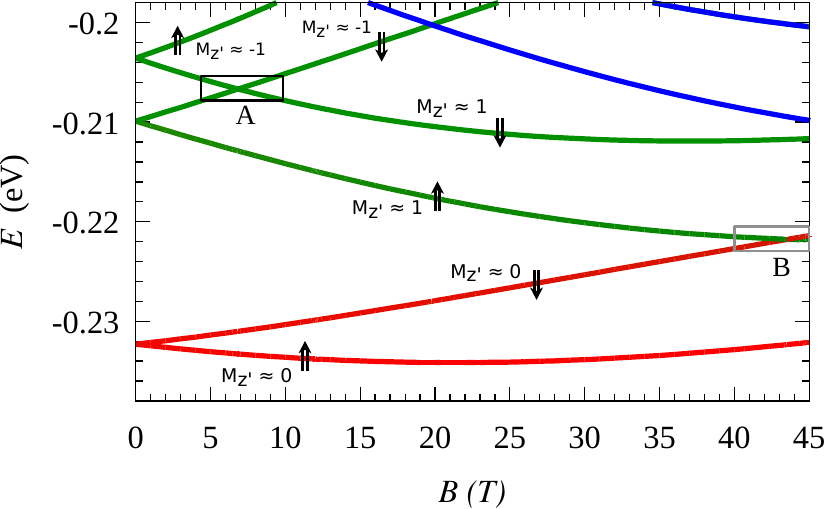}
	\end{center}
	\caption{\label{fig:qd111}\textcolor{gray}({Color online) Magnetic field dependence of the lowest hole energy levels for the lens shaped [111]-oriented QD. } }
\end{figure}
Finally, we investigate the magnetic-field dependence for a lens-shaped QD grown in the [$111$] direction. The simulation results are presented in Fig.~\ref{fig:qd111}.  At $B=0$, the symmetry of the system is $C_{\mathrm{3v}}$, while the axial magnetic field (now oriented along the [$111$] direction) reduces it to $C_{\mathrm{3}}$ (see character tables in Appendix A).  This leads to different selection rules compared to the cases considered previously. We identified the representations of the states:
$s_{1}$ and $s_{2}$ belong to $D_{\mathrm{I}}$, $p_{1}$ and $p_{3}$ to $D_{\mathrm{II}}$, while $p_{2}$ and $p_{4}$ to $D_{\mathrm{III}}$. In consequence, there is no avoided crossing between $p_{2}$ and $p_{3}$ energy branches (see region A in Fig.~\ref{fig:qd111}). Furthermore, $s$- and $p$-type states are decoupled and there is a crossing between their energy levels (a very small anticrossings in the simulation results are numerical artifacts related to the discretization on a rectangular mesh). In contrast to the [$001$]-oriented disk-shaped QD, the crossing between $s_{2}$ and $p_{1}$ energy branches can not be resolved by the axial electric field because it does not change the symmetry of the system. 

\section{Phonon-assisted spin relaxation} 
\label{sec:phonon}

We account for the hole-phonon coupling within the long-wavelength limit.
The phonon-induced spin-flip rate between the states $\ket{\psi_i}$ and $\ket{\psi_j}$  at zero temperature is calculated using the Fermi golden rule
\begin{equation*}
	\gamma_{ij} = \frac{2\pi}{\hbar^2} \sum_{\lambda, \bm{q}} \abs{\matrixel{\psi_j}{\mathcal{V}_\mathrm{int} (\bm{q},\lambda)}{\psi_i}}^2
	\delta \left(\frac{\Delta E_{ij}}{\hbar} - c_{\lambda} q \right),
\end{equation*}
where $\bm{q}$ is a wave vector, $\lambda$ denotes an acoustic phonon branch, $\Delta E_{ij}$ is the energy difference between the states, and $c_\lambda$ is a polarization-dependent speed of sound in GaAs. Finally, $\mathcal{V}_\mathrm{int} (\bm{q},\lambda) = \mathcal{H}^{\mathrm{(ph)}}_{\mathrm{B-P}} (\bm{q},\lambda) + \mathcal{H}^{\mathrm{(ph)}}_{\mathrm{PZ}} (\bm{q},\lambda)$ is a Hamiltonian of hole-phonon interaction containing the couplings via deformation potential (represented by the $8$-band Bir-Pikus Hamiltonian $\mathcal{H}^{\mathrm{(ph)}}_{\mathrm{BP}}$) and via the phonon-induced piezoelectric potential $\mathcal{H}^{\mathrm{(ph)}}_{\mathrm{PZ}}$, both written in terms of phonon modes $(\bm{q},\lambda)$. The details are given in Refs.~\onlinecite{woods04,Climente2013,Roszak2007}. The rotation to [$111$]-oriented coordinate system is performed in a standard way, by transforming strain tensor and invariant matrices.

\begin{figure}[!tb]
	\begin{center}
		\includegraphics[width=3.2in]{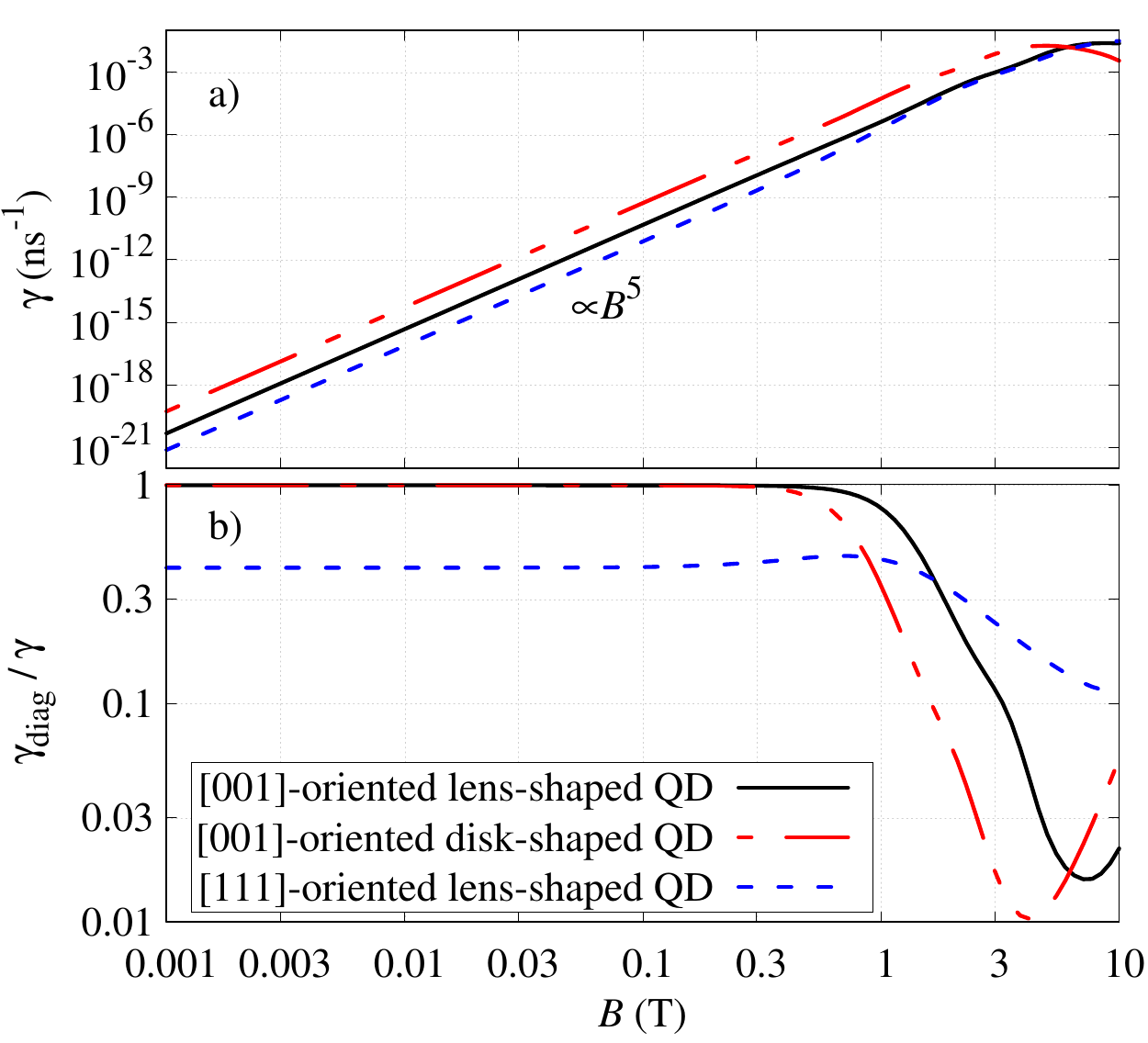}
	\end{center}
	\caption{\label{fig:ph}\textcolor{gray}({Color online) (a) Phonon-assisted spin relaxation rate in the lowest-energy Zeeman doublet as a function of magnetic field for [$001$]- and [$111$]-oriented QDs; (b) ratio of the relaxation rate from spin-admixture mechanisms to the overall relaxation rate. } }
\end{figure}

We calculated phonon-assisted spin relaxation rate from $s_{2}$ to $s_{1}$ state ($\gamma \equiv \gamma_{21}$) for all of the considered QD structures [Fig.~\ref{fig:ph}(a)].  For weak magnetic field, the relaxation rate in the [$111$]-oriented lens-shaped QD is one order of magnitude lower compared to the dot with the same geometry but the [$001$] substrate orientation. This results from the suppression of the $p$-type admixtures ($ M_{z} = \pm 1$) in the heavy-hole components of wave functions in the [$111$]-oriented QD. Furthermore, up to $B \approx 6$~T relaxation in the disk-shaped QD is faster compared to the lens-shaped. This can be related to greater $g$-factor (hence larger transition energy at given value of $B$) in the disk-shaped QD. Finally, at high magnetic field, all of the rates saturate and then decrease, which is caused by the phonon spectral density suppression at high frequency regime. Since for all of the considered structures the $s$-$p$ anticrossing appears at high magnetic field, we do not observe relaxation peaks characteristic for the 2D GaAs QD\cite{Bulaev2005,Bulaev2005b}.
Increase of the spin relaxation time for the [$111$] crystal orientation has been demonstrated (in the regime of strong confinement in growth direction) in a QD defined by harmonic potential in a QW\cite{Lu2005}. In contrast to the QDs in quantum well considered in that paper, the strain field in self-assembled QDs contains nonzero shear components for any crystal orientation. Shear strain enters the hole Hamiltonian with the $d_{\mathrm{v}}$ deformation potential. In consequence, even for the [$001$]-oriented structure, we have a direct strain-related channel of spin admixture, which is important for the spin relaxation process.

We calculated the ratio $\gamma_{\mathrm{diag}}/\gamma$, where $\gamma_{\mathrm{diag}}$ is the transition rate due to all spin-admixture mechanisms, obtained by taking only the spin-diagonal part of $\mathcal{V}_\mathrm{int}$. As shown in Fig.\ref{fig:ph}(b), for the [$001$]-oriented structures the spin-admixture mechanisms dominates at low and moderate magnetic-fields. In contrast, for such fields in [$111$]-oriented structure the spin-admixture part is strongly reduced and direct spin-phonon coupling is relevant in the whole $B$ range. To quantitatively assess the importance of various symmetry contributions to the coupling via spin admixture, we perform an analysis based on spherical harmonics. The spin-admixture (diagonal) part of the Hamiltonian at low magnetic fields is dominated by the piezoelectric potential coupling\cite{Climente2013}. The relevant part of the Hamiltonian can be written as $$\matrixel{\psi_j}{\mathcal{H}^{\mathrm{(ph)}}_{\mathrm{PZ}} (\bm{q},\lambda)}{\psi_i} = M_{\lambda}(\bm{\hat{q}})  \mathcal{F}_{ij}(\bm{q}),$$
where $M_{\lambda}(\bm{\hat{q}})$ is a polarization-dependent geometric factor\cite{grodecka05a}, and $\mathcal{F}_{ij}$ is a form-factor defined by
$$
\mathcal{F}_{ij}(\bm{q}) = \sum_{s=1}^{8} \int_{-\infty}^{\infty} \psi^{*}_{i,s}(\rr) \psi_{j,s}(\rr) e^{i \bm{q} \bm{r}} \mathrm{d} \rr.
$$
The form-factor $\mathcal{F} \equiv \mathcal{F}_{12}$ can be written in spherical coordinates $\mathcal{F}(q,\theta,\phi)$. To obtain values at desired points, we use nonuniform fast Fourier transform in our numerical calculations (NUFFT library)\cite{Lee2005}. Then, the form-factor is expanded in spherical harmonics
$$
\mathcal{F}(q,\theta,\phi) = \sum_{m,l} a_{lm}(q) Y^{m}_l (\theta,\phi),
$$
where
$$
a_{l m}(q) =  \int_{0}^{2 \pi} \mathrm{d} \phi \int_{0}^{\pi} \sin{\theta} \ Y^{m*}_l (\theta,\phi) \mathcal{F}(q,\theta,\phi)  \\ \mathrm{d} \theta.
$$
We calculated $a_{l m}(q)$ coefficients and $W_{l m} = \abs{a_{l m}}^2/\mathcal{N}$, where $$\mathcal{N} = \int_{0}^{2 \pi} \mathrm{d} \phi \int_{0}^{\pi} \sin{\theta} \ \abs{\mathcal{F}(q,\theta,\phi)}^2  \\ \mathrm{d} \theta.$$
\begin{figure}[!tb]
	\begin{center}
		\includegraphics[width=3.2in]{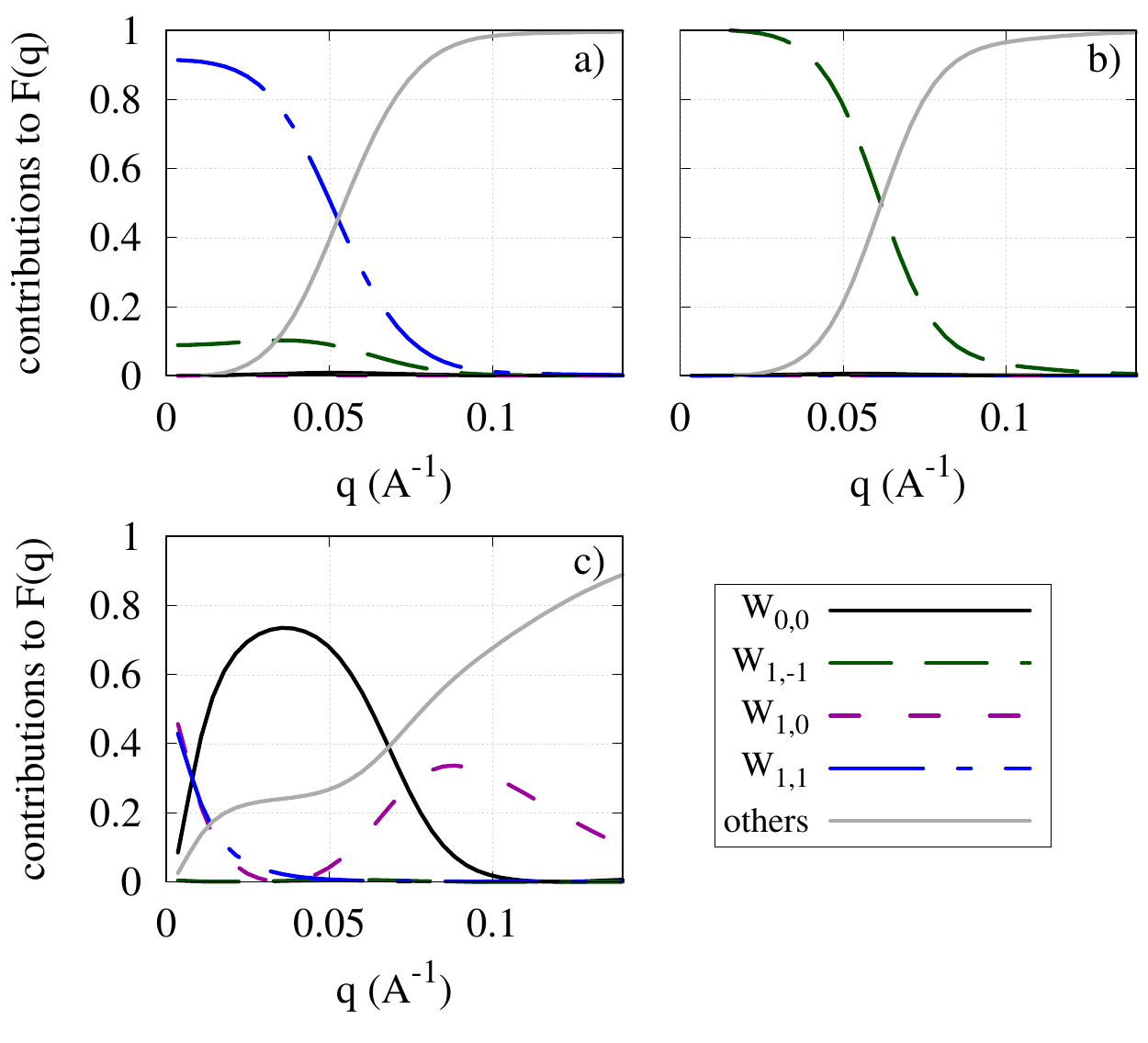}
	\end{center}
	\caption{\label{fig:harm}\textcolor{gray}({Color online) Form-factor $F(q)$ expansion in spherical harmonics for (a) [$001$]-oriented lens- and (b) disk-shaped QD, and (c) [$111$]-oriented lens-shaped QD.} }
\end{figure}
\begin{figure}[!tb]
	\begin{center}
		\includegraphics[width=3.2in]{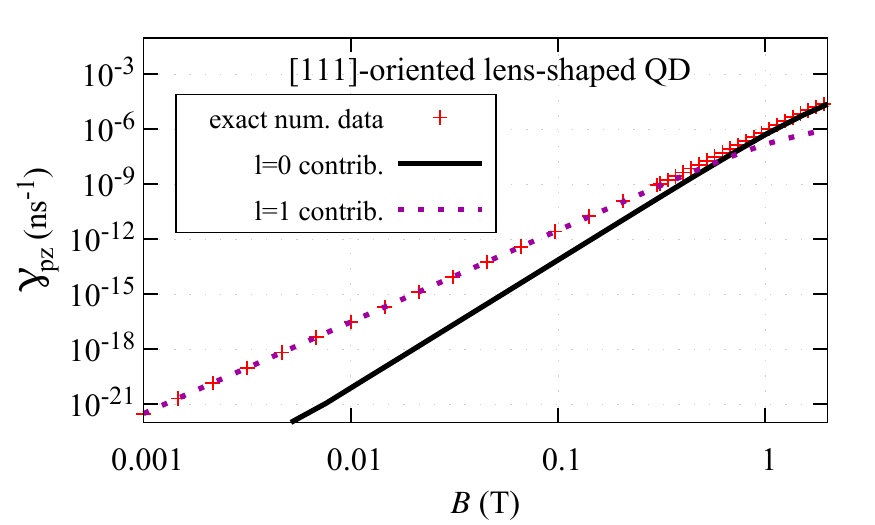}
	\end{center}
	\caption{\label{fig:tun-harm}\textcolor{gray}({Color online) Phonon-assisted spin relaxation rate due to coupling via piezoelectric potential; red points denotes results obtained taking numerically exact $F(q)$, for the black solid line $F(q) \approx a_{0,0} Y^0_0 $, and for the blue dashed line $F(q) \approx a_{1,-1} Y^{-1}_1 + a_{1,0} Y^{0}_1 + a_{1,1} Y^{1}_1$.} }
\end{figure}
Since $\mathcal{F}(\bm{q})$ contains the product of states, the resulting symmetry depends on the relative importance of subbands and their envelopes.  Figs.\ref{fig:harm}(a-c) present the results of $W_{lm}$ for all of the considered structures, where we investigate $s$- and $p$-type components $l=\{0,1\}$ which are dominant at small and moderate $q$. In the case of lens-shaped [$001$]-oriented QD [Fig.\ref{fig:harm}(a)], the main contribution comes from $Y^{1}_{1}$, and considerably smaller from $Y^{-1}_{1}$.
This is consistent to with the results in Sec.~\ref{sec:results}~A, where the coupling between $M_{z} \approx 0$ and $M_{z} \approx 1$ states gives anticrossing between the energy branches. For the disk-shaped structure [Fig.\ref{fig:harm}(b)], the main contribution comes from $Y^{-1}_{1}$. This clearly agrees with the spectrum considered in Sec.~\ref{sec:results}~B, where the selection rules allow coupling between $s_2$ state and nominally $p$ state with $M_{z} \approx -1$, but coupling to $M_{z} \approx 1$ is prohibited.
In the case of [$111$]-oriented structure [Fig.~\ref{fig:harm}(c)], the important contributions at low $q$ come from $Y^{1}_{1}$, $Y^{0}_{1}$ and $Y^{0}_{0}$.
We note, that the states are labeled as $s$- or $p$- type with respect to their dominant envelope symmetry. As discussed in Sec.~\ref{sec:results}~C, the selection rules forbid the coupling of $s_{1}$, $s_{2}$ to nominally $p$-type states with $M_{z} \approx \pm 1$. However, the form-factor contains sum over envelope products from all of the subbands (which have various symmetry, including $p$ type), and their significance depends on $q$. To study the importance of such contributions, we calculate spin relaxation rate due to piezoelectric coupling within two approximations. In the first one, the form-factor contains only the spherical harmonics of $s$-type: $F(q) \approx a_{0,0} Y^0_0 $, and the second one is $F(q) \approx a_{1,-1} Y^{-1}_1 + a_{1,0} Y^{0}_1 + a_{1,1} Y^{1}_1$. As shown in Fig.~\ref{fig:tun-harm}, for magnetic fields up to $B \approx 0.5$~T, the $p$-type contributions dominate. On the other hand, the $s$-type contributions have different magnetic field dependence ($\propto B^5$ vs. $\propto B^7$) and starts to dominate at higher magnetic fields. This explains the change in blue line slope visible in Fig.~\ref{fig:ph}(a).

%
%
%
%
\section{Conclusions}
\label{sec:conclusions}

We have investigated the hole $s$-$p$ coupling related to the spin-orbit interaction for three InAs/GaAs QDs representing symmetry point groups: $C_{\mathrm{2v}}$, $C_{\mathrm{3v}}$ and $D_{\mathrm{2d}}$.
Using the group theory, we have identified irreducible representations of the states and explained the selection rules in the considered QDs. We have shown that in the case of [$001$]-oriented lens shaped QD important contribution to the width of the avoided crossing between $s$- and $p$-shell energy levels comes from the shear strain. Furthermore, we have demonstrated no coupling between nominally $s$- and $p$-type states in the [$111$]-oriented lens shaped QD. We calculated phonon-assisted spin relaxation rates for all of the considered structures and shown that [$111$]-oriented QD offers an order of magnitude longer spin lifetimes. 

\acknowledgments
This work was supported by the
Polish National Science Centre (via Grant No.~2014/13/B/ST3/04603). 
Calculations have been carried out using
resources provided by Wroclaw Centre
for Networking and Supercomputing (\url
{http://wcss.pl}), Grant No.~203.
We would like to thank Pawe{\l} Machnikowski and Micha{\l} Gawe{\l}czyk for their helpful suggestions. We are also grateful to Micha{\l} Gawe{\l}czyk for sharing his implementation of the blur algorithm.

\appendix
\section{Symmetry point groups} 

In this Appendix we present the character tables of the symmetry point groups $C_{\mathrm{2v}}$,  $C_{\mathrm{2}}$, $C_{\mathrm{3v}}$, $C_{\mathrm{3}}$, $D_{\mathrm{2d}}$ and $S_{4}$. In the presence of spin, the double group representations are used. Here $\mathcal{R}$ denotes the rotation of $2 \pi$, while the neutral element $E$ corresponds to the rotation of $4 \pi$\cite{Dresselhaus2010,Bir1974}. In the group $C_{\mathrm{2v}}$, the two-dimensional irreducible representation $D_{1/2}$ contains diagonal matrices $\Gamma^{(D_{1/2})}(E)$, $\Gamma^{(D_{1/2})}(\mathcal{R})$, $\Gamma^{(D_{1/2})}(C_{2})$, but the matrices representing reflections $\sigma_{v}$ have off-diagonal elements. Hence, the reduction $C_{\mathrm{2v}}$ to subgroup $C_{2}$ leaves all $D_{1/2}$ non-diagonal matrices. In conseqence it can be separated into two irreducible representations $D_{A}$, $D_{B}$. 

Within the $8$-band $\kp$ model with envelope function approximation, the eigenstates of the system have a form
\begin{equation*}
\ket{\Psi_{n}} = \sum_{m=1}^{8} \Phi_{n,m} (\rr) \ket{J,J_{z}}_{m},
\end{equation*}
where $\Phi_{n,m}$ is the envelope and $\ket{J,J_{z}}_{m}$ describes the Bloch part (at $\bm{k}=0$) with the total band angular momentum $J$ and its axial projection $J_{z}$. The basis contains: conduction band $\ket{\frac{1}{2},\pm \frac{1}{2}}_{c}$, heavy-hole $\ket{\frac{3}{2},\pm \frac{3}{2}}_{v}$,  light-hole $\ket{\frac{3}{2},\pm \frac{1}{2}}_{v}$, and two split-off subbands $\ket{\frac{1}{2},\pm \frac{1}{2}}_{v}$. To find the irreducible representation of a given state $\ket{\Psi_{n}}$, we performed the projection with operator
$\widehat{P}^{(\alpha)} = \sum_{i}  \chi^{*} (\widehat{R}_{i}) \widehat{R}_{i}$, where $\chi (\widehat{R}_{i}) $ is a character of the representation $\alpha$ for the symmetry operation $\widehat{R}_{i}$\cite{Dresselhaus2010,Piela}. As the envelope part changes slowly in scale of the unit cell, we act with $\widehat{R}_{i}$ on the envelope and Bloch part of the wave functions separately, e.g., the effect of axial rotation $C_{k}$ is $ C_{k} \Phi_{n,m} (\rr) =  \Phi_{n,m} (C^{-1}_{k} \rr)$, and $C_{k} \ket{J,J_{z}} = e^{-i j_{z} 2\pi/k } \ket{J,J_{z}}$. We express the improper rotations $S_{k}$ as $S_{k} = \sigma_{h} C_{k} = \mathcal{I} C_{2} C_{k}$, where $\sigma_{h}$ is reflection in plane perpendicular to the rotation axis and $\mathcal{I}$ is the inversion operator\cite{Bir1974}. The effect of inversion is $\mathcal{I} \ket{\frac{1}{2},\pm \frac{1}{2}}_{c} = \ket{\frac{1}{2},\pm \frac{1}{2}}_{c}$ for the conduction band, and $\mathcal{I} \ket{J,J_{z}}_{v} = -\ket{J,J_{z}}_{v}$ for the valence-band basis states.

\begin{table}[tb]
	\caption{\label{tab:c2v}Character table of $C_{\mathrm{2v}}$ symmetry point group\cite{Bradley1972}.\ }
	\begin{ruledtabular}
		\begin{tabular}{c|ccccc}
			\multirow{2}{*}{$C_{\mathrm{2v}}$} & \multirow{2}{*}{$E$} & \multirow{2}{*}{$\mathcal{R}$} & $C_{2} + $ & $\sigma_{v}(xz)+$ &    $\sigma_{v}(yz)+$ \\ 
			 &  & & $\mathcal{R} C_{2}$ & $\mathcal{R} \sigma_{v}(xz)$ & $\mathcal{R} \sigma_{v}(yz)$ \\
			\hline \\ [-1.5ex]
			$A_{1}$ & 1 & 1 & 1 & 1 & 1\\
			$A_{2}$ & 1 & 1 & 1 & -1 & -1\\
			$B_{1}$ & 1 & 1 & -1 & 1 & -1\\
			$B_{2}$ & 1 & 1 & -1 & -1 & 1 \\
			\hline \\ [-1.5ex]
			$D_{1/2}$ & 2 & -2 & 0 & 0 & 0 
		\end{tabular} 
	\end{ruledtabular}

  \caption{\label{tab:c2}Character table of $C_{2}$ symmetry double point group\cite{Bradley1972}.} 
	\begin{ruledtabular}
		\begin{tabular}{c|cccc}
			$C_{2}$ & $E$ & $C_{2}$ & $\mathcal{R}$ & $\mathcal{R} C_{2}$  \\ 
			\hline \\ [-1.5ex]
			$A_{1}$ & 1 & 1 & 1 & 1 \\
			$B_{1}$ & 1 & -1 & 1 & -1 \\
			\hline \\ [-1.5ex]
			$D_{A}$ & 1 & i & -1 & -i \\
			$D_{B}$ & 1 & -i & -1 & i  
		\end{tabular} 
	\end{ruledtabular}
\end{table} 

\begin{table}[tb]
	\caption{\label{tab:d2d}Character table of $D_{\mathrm{2d}}$  symmetry double point group\cite{Bradley1972}. }
	\begin{ruledtabular}
		\begin{tabular}{c|ccccccc}
			\multirow{2}{*}{$D_{\mathrm{2d}}$} & \multirow{2}{*}{$E$} & \multirow{2}{*}{$\mathcal{R}$} & \multirow{2}{*}{$2S_{4}$} & \multirow{2}{*}{$2\mathcal{R} S_{4}$} &  $C_{2} $ & $2 C'_{2} $ & $2 \sigma_{d}  $ \\ 
			&  &  &  &  &  $\mathcal{R} C_{2} $ & $2 \mathcal{R} C'_{2} $ & $2\mathcal{R} \sigma_{d}$ \\
			\hline \\ [-1.5ex]
			$A_{1}$ & 1 & 1 & 1 & 1 & 1 & 1 &    1\\
			$B_{1}$ & 1 & 1 & $-1$ & $-1$ & 1 & 1 & $-1$\\
			$B_{2}$ & 1 & 1 & $-1$ & $-1$ & 1 & -1 & 1\\
			$E$ & 2 & 2 & 0 & 0 & -2 & 0 & 0 \\
			\hline \\ [-1.5ex]
			$D_{1/2}$ & 2 & $-2$ & $\sqrt{2}$ & $-\sqrt{2}$ & 0 & 0 & 0 \\
			$D'$ & 2 & $-2$ & $-\sqrt{2}$ & $\sqrt{2}$ & 0 & 0 & 0 
		\end{tabular} 
	\end{ruledtabular}

	\caption{\label{tab:s4}Character table of $S_{4}$ symmetry double point group\cite{Bradley1972}.}
	\begin{ruledtabular}
		\begin{tabular}{c|ccccccccc}
			$S_{4}$ & $E$ & $S_{4}$ & $C_{2}$ & $S^{3}_{4}$  & $\mathcal{R}$ & $\mathcal{R} S_{4}$ & $\mathcal{R} C_{2}$ & $\mathcal{R} S^{3}_{4}$ \\ 
			\hline \\ [-1.5ex]
			$A$ & 1 & 1 & 1 & 1 & 1 & 1 & 1 & 1\\
			$B$ & 1 & $-1$ & 1 & $-1$ & 1 & $-1$ & 1 & $-1$\\
			$E_{1}$ & 1 & $i$ & $-1$ & $-i$ & 1 & $i$ & $-1$ & $-i$\\
			$E_{2}$ & 1 & $-i$ & $-1$ & $i$ & 1 & $-i$ & $-1$ & $i$\\ [1ex]
			\hline \\ [-1.5ex]
			$D_{\mathrm{I}}$ & 1 & $\frac{-1+i}{\sqrt{2}}$ & $-i$ & $\frac{1+i}{\sqrt{2}}$ & $-1$ & $\frac{1-i}{\sqrt{2}}$ & $i$ & $\frac{-1-i}{\sqrt{2}}$ \\ [1ex]
			$D_{\mathrm{II}}$ & 1 & $\frac{-1-i}{\sqrt{2}}$ & $i$ & $\frac{1-i}{\sqrt{2}}$ & $-1$ & $\frac{1+i}{\sqrt{2}}$ & $-i$ & $\frac{-1+i}{\sqrt{2}}$ \\ [1ex] 
			$D_{\mathrm{III}}$ & 1 & $\frac{1-i}{\sqrt{2}}$ & $-i$ & $\frac{-1-i}{\sqrt{2}}$ & $-1$ & $\frac{-1+i}{\sqrt{2}}$ & $i$ & $\frac{1+i}{\sqrt{2}}$ \\  [1ex] 
			$D_{\mathrm{IV}}$ & 1 & $\frac{1+i}{\sqrt{2}}$ & $i$ & $\frac{-1+i}{\sqrt{2}}$ & $-1$ & $\frac{-1-i}{\sqrt{2}}$ & $-i$ & $\frac{1-i}{\sqrt{2}}$ \\  [1ex]
		\end{tabular} 
	\end{ruledtabular}

\end{table} 

\begin{table}[tb]

	\caption{\label{tab:c3v}Character table of $C_{\mathrm{3v}}$ symmetry double point group\cite{Bradley1972}.  }
	\begin{ruledtabular}
		\begin{tabular}{c|cccccc}
			$C_{\mathrm{3v}}$ & $E$ & $\mathcal{R}$ & $2C^{2}_{3}$ & $2 \mathcal{R} C^{2}_{3}$ &  $3 \sigma_{v}$ & $3 \mathcal{R} \sigma_{v}$  \\ 
			\hline \\ [-1.5ex]
			$A_{1}$ & 1 & 1 & 1 & 1 & 1 & 1\\
			$A_{2}$ & 1 & 1 & 1 & 1 & -1 & -1\\
			$E$ & 2 & 2 & -1 & -1 & 0 & 0 \\
			\hline \\ [-1.5ex]
			$D_{1/2}$ & 2 & -2 & 1 & -1 & 0  & 0\\
			$D'$ & 1 & -1 & -1 & 1 & i  & -i \\
			$D''$ & 1 & -1 & -1 & 1 & -i  & i 
		\end{tabular} 
	\end{ruledtabular}

	\caption{\label{tab:c3}Character table of $C_{3}$ symmetry double point group\cite{Bradley1972}. }
  \begin{ruledtabular}
	  \begin{tabular}{c|cccccc}
			$C_{3}$ & $E$ & $C_{3}$ & $C^{2}_{3}$ & $\mathcal{R}$ &  $\mathcal{R} C_{3}$ & $\mathcal{R} C^{2}_{3}$ \\ 
			\hline \\ [-1.5ex]
			$A_{1}$ & 1 & 1 & 1 & 1 & 1 & 1\\
			$B_{1}$ & 1 & $e^{2i \pi /3}$ & $e^{4i \pi /3}$ & 1 & $e^{2i \pi /3}$ & $e^{4i \pi /3}$\\
			$B_{2}$ & 1 & $-e^{i \pi /3}$ & $e^{2i \pi /3}$ & 1 & $-e^{i \pi /3}$ & $e^{2i \pi /3}$\\
			\hline \\ [-1.5ex]
			$D_{\mathrm{I}}$ & 1 & $-1$ & $1$ & $-1$ & $1$ & $-1$ \\
			$D_{\mathrm{II}}$ & 1 & $e^{i \pi /3}$ & $e^{i 2 \pi /3}$ & $-1$ & $-e^{i \pi /3}$  & $-e^{i 2 \pi /3}$\\
			$D_{\mathrm{III}}$ & 1 & $-e^{ 2 i \pi /3}$ & $e^{ 4 i \pi /3}$ & $-1$ & $e^{ 2 i \pi /3}$  & $-e^{ 4 i \pi /3}$
		\end{tabular} 
	\end{ruledtabular}

\end{table} 

\section{Effective model}

In this part we describe the effective model that can be used to interpret the simulation results. We utilize the Fock-Darwin model supplemented by additional terms representing system anisotropy as well as the spin-orbit coupling\cite{Manaselyan09,Siranush13,Gawarecki2018a}. 

In the axial approximation, the states in a QD can be characterized according to their axial projection of the envelope angular momentum $M_{z}$, where the $s$ shell contains states with $M_{z} = 0$ and the $p$ shell with $M_{z} = \pm 1$.  In fact, $p$-type states can be mixed due to anisotropy related to the piezoelectric potential, dot elongation and other possible effects. Due to the dominant heavy-hole components of the considered states, their axial projections  of band angular momenta ($\Uparrow,\Downarrow$) can be approximated by $\langle J_{z} \rangle \approx \pm 3/2$. Furthermore, the spin-orbit coupling distinguishes the mutual alignment of the envelope and the band angular momenta as well as it can mix $s$- and the $p$-shell states. We express the Hamiltonian in the basis $\ket{M_{z} J_{z}} = \ket{M_{z}} \otimes \ket{J_{z}}$ and consider $s$ and $p$ shells $\qty{\ket{0 \Uparrow}, \ket{1 \Uparrow}, \ket{-1 \Uparrow}, \ket{0 \Downarrow}, \ket{1 \Downarrow}, \ket{-1 \Downarrow} }$. The effective Hamiltonian reads
\begin{align*} \label{eff_ham}
		H_\mathrm{eff} &= E_{\mathrm{s}} \ketbra{0}{0} \otimes \mathbb{I}_{2} + E_{\mathrm{p}} \qty( \ketbra{1}{1} + \ketbra{-1}{-1} ) \otimes \mathbb{I}_{2} \\ 
		&\phantom{=}                  +  V_{\mathrm{a}} \qty( \ketbra{1}{-1} + \ketbra{-1}{1} )\otimes \mathbb{I}_{2}  + \frac{1}{\hbar} W B_{z} L_{z} \otimes \mathbb{I}_2 \\
		&\phantom{=}
		+  \frac{1}{2}  \mu_{B} \qty [ g_{\mathrm{s}} \ketbra{0}{0}
		+  g_{\mathrm{p}} (\ketbra{1}{1} + \ketbra{-1}{-1}) ] B_{z}  \otimes \sigma_{z}   \\
		&\phantom{=} + \frac{1}{2 \hbar}  V^{\mathrm{(so)}}_{\mathrm{pp}} L_{z} \otimes \sigma_{z}  \\
		&\phantom{=} + V^{\mathrm{(so)}}_{\mathrm{sp}} \qty ( \ketbra{0}{-1} \otimes \ketbra{\Uparrow}{\Downarrow}  + \ketbra{-1}{0} \otimes \ketbra{\Downarrow}{\Uparrow} ) \\
		&\phantom{=} - V^{\mathrm{(so)}}_{\mathrm{sp}} \qty ( \ketbra{0}{1} \otimes \ketbra{\Downarrow}{\Uparrow} + \ketbra{1}{0} \otimes \ketbra{\Uparrow}{\Downarrow} ) \\
		&\phantom{=} + \alpha_{\mathrm{s}} B^{2}_{z} \ketbra{0}{0} \otimes \mathbb{I}_2  \\ 
		&\phantom{=} + \alpha_{\mathrm{p}} B^{2}_{z} (\ketbra{1}{1} + \ketbra{-1}{-1}) \otimes \mathbb{I}_2, 
\end{align*}
where $E_{\mathrm{s}}$, $E_{\mathrm{p}}$ are the bare energies ($B=0$, axial approximation, SOC neglected) of the $s$- and $p$-type states respectively,  $\mathbb{I}_2$ is the unit operator in the band angular momentum formal subsystem,  $V_{\mathrm{a}}$ is a parameter accounting for the anisotropy, $L_{z}$ is the operator of the $z$ component of the envelope angular momentum, $g_{\mathrm{s}}$ and $g_{\mathrm{p}}$ are $g$-factors in $s$- and $p$-shell respectively, $\sigma_{i}$ are the Pauli matrices,
$V^{\mathrm{(so)}}_{\mathrm{pp}}$ describes the spin-orbit coupling for the $p$-states, $V^{\mathrm{(so)}}_{\mathrm{sp}}$ is a parameter related to the coupling between  $s$ and $p$ states, finally $\alpha_{\mathrm{s}}$ and $\alpha_{\mathrm{p}}$ account for the diamagnetic shift. We neglect the coupling of $\ket{0 \Uparrow}$ to $\ket{1 \Downarrow}$, and $\ket{0 \Downarrow}$ to $\ket{-1 \Uparrow}$ because they are not represented by any avoided crossing in the considered spectrum. The effective Hamiltonian can be then written in matrix block form
\begin{equation*} \label{eff_ham}
	\begin{split}
		H_\mathrm{eff} = 
		\mqty(
		\mathcal{H}_{\mathrm{env}} + \frac{1}{2} \mathcal{H}_{\mathrm{1}}& \mathcal{H}_{\mathrm{2}} \\
		\mathcal{H}^{\dagger}_{\mathrm{2}} & \mathcal{H}_{\mathrm{env}} - \frac{1}{2}  \mathcal{H}_{\mathrm{1}} ),
	\end{split}
\end{equation*}
where
\begin{align*} 
	\begin{split}
		\mathcal{H}_{\mathrm{env}} =& 
		\mqty(
		E_{\mathrm{s}}  & 0 & 0 \\
		0 & E_{\mathrm{p}} + W B_{z} & V_{\mathrm{a}} \\
		0 & V_{\mathrm{a}} & E_{\mathrm{p}} - W B_{z} )  \\
		& + \mqty(
		\alpha_{\mathrm{s}} B^{2}_{z} & 0 & 0 \\
		0 & \alpha_{\mathrm{p}} B^{2}_{z} & 0 \\
		0 & 0 & \alpha_{\mathrm{p}} B^{2}_{z}),
	\end{split}
\end{align*}
\begin{align*} 
	\begin{split}
		\mathcal{H}_{\mathrm{1}} =& 
		\mqty(
		\mu_{B} g_{\mathrm{s}} B_{z}  & 0 & 0 \\
		0 &  \mu_{B} g_{\mathrm{p}} B_{z} +  V^{\mathrm{(so)}}_{\mathrm{pp}}  & 0 \\
		0 & 0 &  \mu_{B} g_{\mathrm{p}} B_{z} - V^{\mathrm{(so)}}_{\mathrm{pp}} ),
	\end{split}
	\\
	\begin{split}
		\mathcal{H}_{\mathrm{2}} =& 
		\mqty(
		0 & 0 & V^{\mathrm{(so)}}_{\mathrm{sp}} \\
		-V^{\mathrm{(so)}}_{\mathrm{sp}} & 0 & 0 \\
		0 & 0 & 0 ).
	\end{split}
\end{align*}

We fitted the simulation data from Fig.~\ref{fig:mag_h} with the effective model and obtained the following parameters: $E_{\mathrm{s}} = -229.14$~meV, $E_{\mathrm{p}} = -203.75$~meV, $V_{\mathrm{a}} = 0.33328$~meV, $W = -0.46764$~meV/T, $g_{\mathrm{s}} = -5.5745$, $g_{\mathrm{p}} = -0.11141$, $V^{\mathrm{(so)}}_{\mathrm{pp}} = -8.0707$~meV, $V^{\mathrm{(so)}}_{\mathrm{sp}} = 123.13$~$\mathrm{\mu eV}$,   $\alpha_{\mathrm{s}} = 3.0834$~$\mathrm{\mu eV/T^{2}}$, and $\alpha_{\mathrm{p}} = 5.0050$~$\mathrm{\mu eV/T^{2}}$. Such parameter set gives energies in a good agreement with these obtained from the $8$-band $\kp$ model.


\bibliographystyle{prsty}
\bibliography{abbr,./library.bib}

\end{document}